\documentclass[12pt]{iopart}

\usepackage{iopams}
\usepackage{graphicx}
\usepackage{cite}
\usepackage{caption}
\usepackage{float}
\usepackage{color}
\usepackage{ulem}

\begin{document}

\title{Blind quantum computation for a user who only performs single-qubit gates}

\author{Qin Li}
\address{School of Computer Science, Xiangtan
University, Xiangtan 411105, China}
\ead{liqin@xtu.edu.cn}
\author{Chengdong Liu, Yu Peng}
\address{China General Technology Research Institute, Beijing 100091, China}
\author{Fang Yu}
\address{Faculty of Informatics, Masaryk University, Brno, Czech Republic\\Department of Computer Science, Jinan University, Guangzhou 510632, China}
\author{Cai Zhang}
\address{College of Mathematics and Informatics, South China Agricultural University,
Guangzhou 510642, China}

\vspace{10pt}

\begin{abstract}
Blind quantum computation (BQC) allows a user who has limited quantum capability to complete a quantum computational task with the aid of a remote quantum server, such that the user's input, output, and even the algorithm can be kept hidden from the server. Up to now, there are mainly two models of BQC. One is that the client just needs the ability to prepare single qubits initiated by Broadbent, Fitzsimons, and Kashefi, and the other is that the client only needs perform single-qubit measurements first given by Morimae. In this paper, we put forward a new model of BQC in which a user only requires implementing a few single-qubit gates. We also propose a specific BQC protocol where a user only needs to implement two kinds of single-qubit gates to show the feasibility of the presented model. This circuit model is quite flexible since various users with the ability to perform different single-qubit gates may all have the chance to achieve BQC. Furthermore, compared with the other two models, it may be more suitable for practical implementation in some experimental setups such as trapped ions and superconducting systems since the single-qubit gates are the most exact operations in such systems.
\end{abstract}

%
%

%
%
%
\section{Introduction}\label{section1}
 Quantum computation based on quantum mechanical principles can provide a considerable advantage over its classical counterpart in solving some problems \cite{NC10:QCI}. Here gives three famous examples. Quantum mechanical systems which are rather difficult to be simulated by classical computers can be efficiently simulated by quantum computers \cite{Feynman:SM:IJTP82}; particularly, Shor's algorithms for factorizing big integers and solving discrete logarithm problems obtain an exponential speed advantage over their best-known classical algorithms \cite{Shor:AQC:FOCS94}, and Grover's algorithm for searching offers a quadratic speedup over its best-known classical counterparts \cite{Grover:QS:PRL97}.

However, the experimental realization of quantum computers is still rather challenging. Although scientists and engineers are strived to find various appropriate physical systems to build quantum computers, all of them just allows for simple operations on a few qubits until now. Even if recently IBM and Google separately announced they have bulit quantum computers which can control more than fifty qubits, to build large-scale universal quantum computers is still a long way to go. The first generation of quantum computers will be most likely to be the ``cloud'' style \cite{MF:BQC:PRA13}; that is,
a few of costly quantum servers held by academia, corporations, or governments will be remotely accessed by users with limited computational resources or power. In such a situation, suppose that a user Alice needs to delegate her computational problems that cannot be efficiently done on classical computers to a quantum server Bob, and she also wants to keep all the data including input, output, and algorithm private. Then BQC comes into being as a solution.

In 2005, Childs proposed the first BQC protocol where the user Alice is required to own quantum memory, prepare a state $|0\rangle$, and be able to implement Pauli X gate, Pauli Z gate, and the SWAP gate \cite{Childs:BQC:QIC05}. Then Arrighi and Salvail devised a BQC protocol in which Alice just needs to prepare and measure entangled states \cite{AS:BQC:IJQIC06}. However, this protocol is not a universal one since it only allows for calculating some particular functions. Until 2009, Broadbent, Fitzsimons, and Kashefi presented the first universal BQC protocol (namely the famous BFK protocol) where Alice only requires preparing single-qubit states \cite{BFK:UBQC:FOCS09}. Furthermore, the protocol has already been experimentally demonstrated by performing a series of blind computations on four-quit blind cluster states \cite{BEBFZP:DBQC:S12}.  In 2013, Morimae proposed another type of BQC protocol where Alice only make single-qubit measurements \cite{MF:BQC:PRA13} since in certain experimental setups, such as optical systems, the measurement of a state is easier than the preparation of a state.  Besides, various BQC protocols also have been devised to be as practical as possible \cite{ABE:BQC:ICS10,DKL:BQC:PRL12,MF:BQC:NC12,GMMR:BQC:PRL13,MPFF:BQC:PRL13,Qin:BQC:PRA14,perez2015iterated,chien2015fault,fujii2017verifiable,fit2017private,gheorghiu2018simple,sata2019arbitrable}, such as making Alice as classical as possible or tolerating more faults, or have been constructed to own some special properties \cite{sheng2015deterministic,takeuchi2016blind,sheng2018blind,morimae2014verification,hayashi2015verifiable,morimae2016measurement,fitzsimons2017unconditionally,li2018blind}, such as adapting to noise environments or realizing verification. Recently, Ferracin et al. proposed a very ingenious mesothetic verification protocol where the verifier Alice requires an $n$-qubit memory and the ability to execute single-qubit gates and also posed an interesting question that whether a mesothetic protocol can be devised that only requires single-qubit gates and single-qubit memory for the verifier \cite{ferracin2019accrediting}.

In this paper, we propose a novel BQC model where the user Alice only needs performing single-qubit gates and give a specific universal BQC protocol in which Alice is just capable of implementing two kinds of single-qubit operations on the received qubits based on the famous BFK protocol \cite{BFK:UBQC:FOCS09} to show its feasibility. The proposed model is rather flexible which allows users with the ability to implement different single-qubit gates to achieve BQC and thus will extend its application. Moreover, we also answer the question raised by Ferracin et al. in Ref. \cite{ferracin2019accrediting} to some extent. The most important is that it is quite suitable for practical realization due to empirical observation \cite{ferracin2019accrediting} that the single-qubit gates are the most exact operation in some experimental setups such as trapped ions \cite{harty2016high} and superconducting systems \cite{barends2014superconducting}.

\section{A new BQC model where Alice only implements single-qubit gates}

There are some ways to realize universal quantum computation. For instance, Dodd et al. proved that a general entangling Hamiltonian and single qubit gates would suffice to produce universal quantum computation \cite{dodd2002universal}. Parra-Rodriguez showed an entangling Hamiltonian on all the time and sending single-qubit pulses is an efficient and effective way of performing universal computation \cite{parra-rodriguez2020digital}. They all may be considered to realize universal BQC. In this paper we will adopt the way of using a specific set of universal gates. As the Ref. \cite{BMPRV:BQC:PRL13} showed that the gate set that consists of two single-qubit gates $H$, $\sigma_z^{1/4}$, and a two-qubit gate $CNOT$ is universal for quantum computation. Thus the main work of such a BQC model where Alice only performs single-qubit gates is to complete these three gates with the aid of a quantum server Bob and the most difficult point is to achieve the two-qubit $CNOT$ gate. In the following, we show that Alice who is only capable of implementing a few single-qubit gates can achieve such a universal set with the help of Bob by considering two cases.

\begin{figure}
\centering
  \includegraphics[scale=0.6]{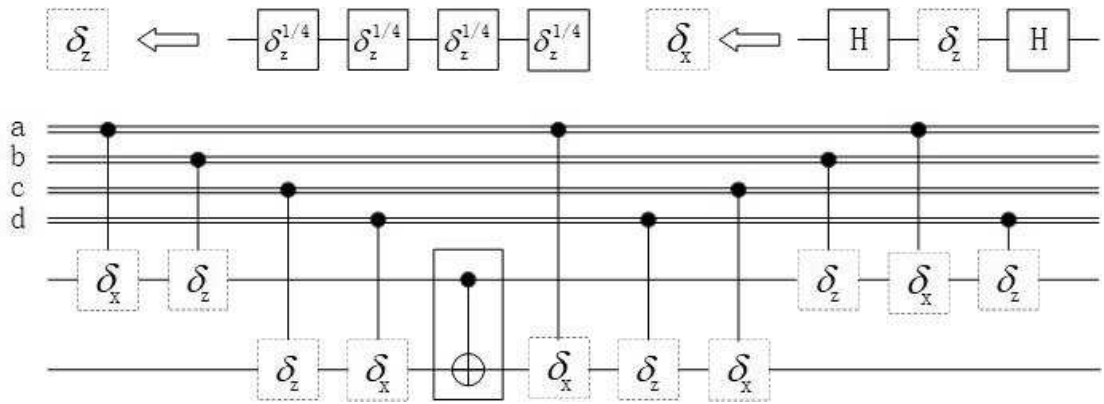}
  \caption{The figure illustrates that Alice who only has the ability to implement gates $H$ and $\sigma_z^{1/4}$ can achieve the $CNOT$ operation with the aid of Bob. The solid square contains the gate that Alice can perform directly by herself, the dashed square contains the gate that can be achieved by Alice through the combined use of gates $H$ and $\sigma_z^{1/4}$, and the solid rectangle contains the two-qubit $CNOT$ operation that Bob does for Alice.}\label{fig:1}
\end{figure}

For simplicity, the first case is that we suppose Alice only can implement the gates $H$ and $\sigma_z^{1/4}$. Then the Pauli $Z$ gate $\sigma_z$ can be achieved by the repeated use of $\sigma_z^{1/4}$ since $\sigma_z=\sigma_z^{1/4}\times\sigma_z^{1/4}\times\sigma_z^{1/4}\times\sigma_z^{1/4}$. Similarly, the Pauli $X$ gate $\sigma_x$ also can be done by the combined use of $H$ and $\sigma_z^{1/4}$ due to $\sigma_x=H(\sigma_z^{1/4})^4H$. Alice can use the Pauli $X$ and the Pauli $Z$ gates to achieve the $CNOT$ operation assisted by the server Bob in Ref. \cite{Childs:BQC:QIC05}, thus Alice also can finish the $CNOT$ operation by using some $H$ and $\sigma_z^{1/4}$ gates with the help of quantum Bob. In a BQC model, as the qubits to be operated should be kept hidden from the server Bob, Alice needs to randomly choose four classical bits $a$, $b$, $c$, and $d$ for encrypting the two input qubits of the $CNOT$ gate. After she performs $((\sigma_z^{1/4})^4)^b(H(\sigma_z^{1/4})^4H)^a$ on the first qubit and $((\sigma_z^{1/4})^4)^d(H(\sigma_z^{1/4})^4H)^c$ on the second qubit, she sends them to Bob and asks Bob to implement the $CNOT$ operation. Alice can obtain the results returned by Bob by implementing the corresponding operations as shown in Fig. \ref{fig:1}. In such a BQC model, the server Bob cannot obtain anything about Alice's private data except the number of the $CNOT$ gates that Alice asks Bob to perform due to the use of quantum one-time pad. Even the number of the used $CNOT$ gates can be hidden if in some positions Alice asks Bob to implement $CNOT$ gates on some trap qubits.

\begin{figure}
\centering
  \includegraphics[width=1.0\linewidth]{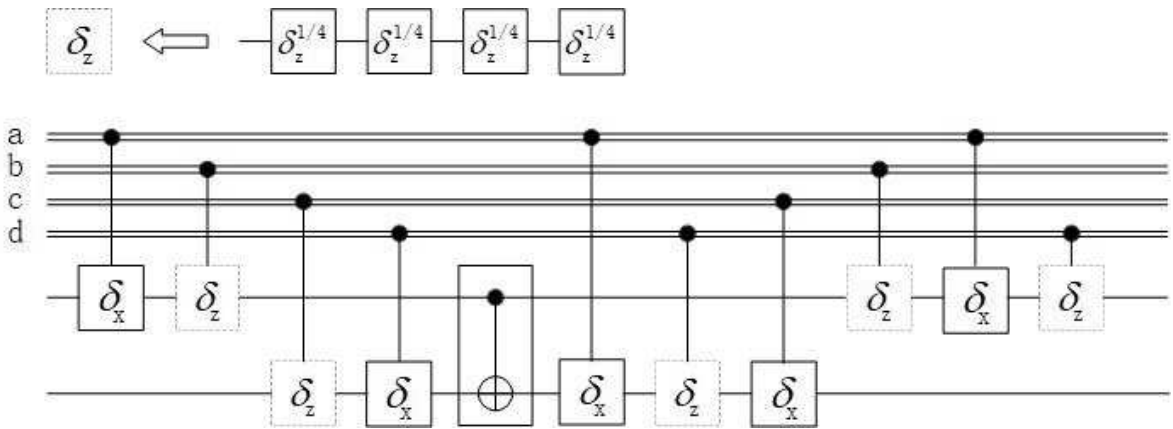}
  \caption{The figure illustrates that Alice who just can perform gates $X$ and $\sigma_z^{1/4}$ can achieve the $CNOT$ operation with the help of Bob. The solid square contains the gate that Alice can perform directly by herself, the dashed square contains the gate that can be achieved by Alice through the combined use of gates $\sigma_x$ and $\sigma_z^{1/4}$, and the solid rectangle contains the two-qubit $CNOT$ operation that Bob does for Alice.}\label{fig:2}
\end{figure}

\begin{figure}
\centering
  \includegraphics[width=0.7\linewidth]{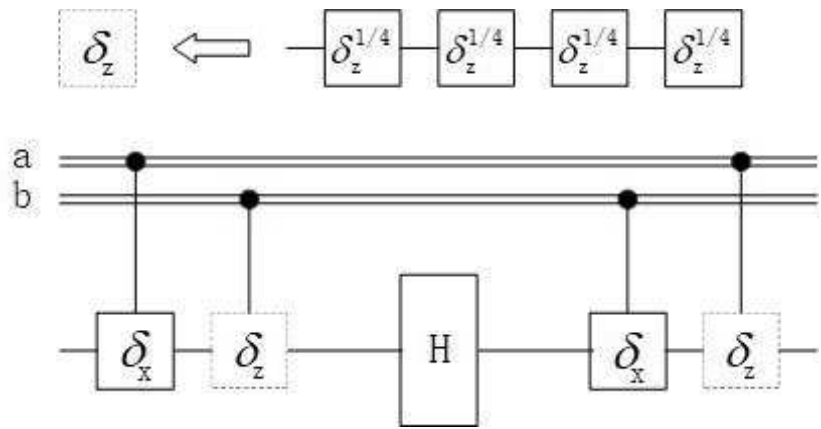}
  \caption{The figure illustrates that Alice who is only capable of performing gates $\sigma_x$ and $\sigma_z^{1/4}$ can achieve the $H$ gate with the help of Bob. The solid square contains the gate that Alice can perform directly by herself, the dashed square contains the gate that can be achieved by Alice through the combined use of gates $\sigma_x$ and $\sigma_z^{1/4}$, and the solid rectangle contains the $H$ gate that Bob does for Alice.}\label{fig:3}
\end{figure}

The second case is that we suppose Alice only has the ability to perform the Pauli $X$ gate $\sigma_x$ and the $\pi/8$ gate $\sigma_z^{1/4}$. Since the $CNOT$ operation can be securely done by using $\sigma_x$ and $\sigma_z$ with Bob's help \cite{Childs:BQC:QIC05}, it is not difficult to infer that the $CNOT$ operation can be done by Alice with the aid of Bob due to $\sigma_z=\sigma_z^{1/4}\times\sigma_z^{1/4}\times\sigma_z^{1/4}\times\sigma_z^{1/4}$, as shown in Fig. \ref{fig:2}.
Similarly, the $H$ gate also can be securely performed by the use of $\sigma_x$ and $\sigma_z^{1/4}$ with the help of Bob since Alice can encrypt her input with gates $\sigma_x$ and $\sigma_z^{1/4}$ and ask Bob to implement the $H$ gate for her, and the specific method can be found in Fig. \ref{fig:3}. In such a BQC model, as the input is encrypted, the server Bob cannot obtain anything about Alice's private data except that the number of the $CNOT$ gates and that of the $H$ gates that Alice asks Bob to perform because of the use of quantum one-time pad. Similarly, the number of the used $CNOT$ or $H$ gates can be kept private if in some positions Alice asks Bob to implement  $CNOT$ or $H$ gates on some trap qubits.

\section{An example in the proposed model based on the BFK protocol}
We first simply review the BFK protocol and then construct a specific BQC protocol in the proposed model where Alice only does a few single-qubit operations. In the proposed protocol, Alice only needs the ability to implement the Hadamard gate $H$ and the $\pi/8$ gate $\sigma_z^{1/4}$. Furthermore, Alice can achieve universal quantum computation with the help of a quantum server Bob.

\subsection{Review of the BFK protocol \cite{BFK:UBQC:FOCS09}}

Suppose that the client Alice has in mind the quantum computation on the $m$-qubit graph state corresponding to the graph $G$. The specific operation she intends to carry out is to measure the $i$th qubit in the basis $\{|\pm\phi_i\rangle=\frac{1}{\sqrt2}(|0\rangle\pm e^{i\phi_i}|1\rangle)\}$, where $\phi_i\in S\equiv\{(k\pi)/4|k=0,1,\ldots,7\}$. Then the single-server BQC protocol can be briefly described as follows.

(S1) Alice prepares $m$ qubits and sends them to the server Bob. The state of each qubit is $|\theta_i\rangle=\frac{1}{\sqrt2}(|0\rangle+e^{i\theta_i}|1\rangle) (i=1,2,\ldots,m)$, where $\theta_i$ is uniformly chosen from the set $S$. Note that if Alice inserts extra $l$ trap qubits, the states of which are one of $\{|0\rangle,|1\rangle\}$, she can verify the computation afterwards \cite{fitzsimons2017unconditionally}.

(S2) Alice asks Bob to generate a brickwork state according to the graph $G$ specified by her.

(S3) Bob produces the brickwork state $|G(\theta)\rangle$ by applying controlled-$Z$ gates on the received qubits based on the graph $G$.

(S4) In each position, Alice randomly chooses $r_i$ and computes $\delta_i=(\theta_i+\phi_i'+r_i\pi)$ mod $2\pi$, where $\phi_i'$ is obtained according to the previous measurements and $\phi_i$, and then $\delta_i$ is sent to Bob if Alice needs Bob to measure the $i$-th qubit of $|G(\theta)\rangle$.

(S5) Bob performs a measurement on the $i$-th qubit in the basis $\{|\pm\delta_i\rangle\}$ for each received qubit and informs Alice about the measurement result.

\subsection{The proposed protocol}
Protocol 1: Suppose the client Alice just has the ability to apply single-qubit gates $H$ and $\sigma_z^{1/4}$ and she also has the same goal like the BFK protocol. She wants to finish the computation on the $m$-qubit graph state corresponding to the graph $G$. The specific steps are in the following.

(A1) Alice asks a quantum server Bob to initiate the task of BQC.

(A2) Bob sends to Alice $m+k+l$ qubits in the computational-based state $|0\rangle$, where $m$ qubits will be used for computation as the original BFK protocol, $k$ qubits will be taken as decoy qubits used for Alice and Bob to detect whether the transmitted states are changed by an outside attacker, and the rest $l$ qubits will be considered as trap qubits used for Alice to verify the computation performed by Bob later.

(A3) When each qubit arrives, Alice performs one of the following three operations: 1) if it is chosen to be used for computation, Alice first performs $H$ gate on it and then randomly applies the gate $\sigma_z^{1/4}$ on it for $n$ times in order to make the state of qubit be $\frac{1}{\sqrt2}(|0\rangle\pm e^{\frac{n\pi}{4}i}|1\rangle)$, where $n$ is uniformly chosen from $\{0,1,\cdots,7\}$;  2) if it is used as a decoy qubit, Alice performs the corresponding single-qubit gates to make the state of the decoy qubit be one of $\{|0\rangle,|1\rangle,|+\rangle,|-\rangle\}$, where $|0\rangle=|0\rangle, |1\rangle=H(\sigma_z^{1/4})^4H|0\rangle, |+\rangle=H|0\rangle,$ and $|-\rangle=(\sigma_z^{1/4})^4H|0\rangle$; and 3) if it is considered as a trap qubit, Alice does nothing or performs $H(\sigma_z^{1/4})^4H$ on it to produce $|0\rangle$ or $|1\rangle$. All of these qubits will be sent to Bob after Alice generates her expected states.

(A4) Alice reveals which qubits are decoy qubits and Bob measures these $k$ trap qubits in the basis $\{|0\rangle,|1\rangle\}$ or $\{|+\rangle,|-\rangle\}$ and reveal the results. Alice and Bob compare the results like the BB84 protocol. If the server Bob can always publish correct results when he chose the right bases, Alice and Bob think that there was no attack during the transmission in step (A2) and step (A3) and Alice will use the rest $m+l$ qubits for computation and verification; otherwise the protocol aborts.

The subsequent steps (A5)-(A8) are the same as the steps (S2)-(S5) of the BFK protocol in Ref. \cite{BFK:UBQC:FOCS09} or the modified verifiable BFK protocol in Ref.\cite{fitzsimons2017unconditionally}.

The main difference between the proposed protocol and the BFK protocol is that Alice performs some sing-qubit operations on the quantum states sent by the server Bob instead of generating single-qubit states by herself. Since Alice asks Bob to measure $k$ qubits in random bases $\{|0\rangle,|1\rangle\}$ or $\{|+\rangle,|-\rangle\}$ and reveal the results, an outside attacker will be discovered with nonzero probability if the transmitted states in step (A2) or step (A3) were not the ones as required and the probability will be reduced to 1 as $k$ becomes big enough. For example, suppose a third party replaces $|0\rangle$ with the Bell state $\frac{|00\rangle+|11\rangle}{\sqrt 2}$ and sends one qubit to Alice while keeping the other qubit by himself in step (A2). If it is considered as a decoy qubit, Alice randomly performs $I$, $H$, $(\sigma_z^{1/4})^4H$, or $H(\sigma_z^{1/4})^4H$ on it and the state of the two-qubit system will be $\frac{|00\rangle+|11\rangle}{\sqrt 2}$, $\frac{|+0\rangle+|-1\rangle}{\sqrt 2}$, $\frac{|-0\rangle+|+1\rangle}{\sqrt 2}$, or  $\frac{|10\rangle+|01\rangle}{\sqrt 2}$. No matter the attacker chose either $\{|0\rangle,|1\rangle\}$ or $\{|+\rangle,|-\rangle\}$ to measure the qubits kept by him, he may be detected with probability 1/2 even if he chose the same bases as Bob did. Then he will be discovered with probability $1-(\frac{1}{2})^k$ which is reduced to 1 as $k$ becomes big enough. 

Note that Bob also can choose not to prepare the states as required and may escape check in step (A4), but it is not helpful for him to obtain Alice's private information since he cannot distinguish which operations Alice did in step (A3). Just like the BFK protocol, Bob can replace all the single-qubit states sent by Alice with other states, but it is useless for Bob to obtain Alice's information. Thus if the security check is passed in step (A4), the proposed protocol can obtain the similar security level as the BFK protocol due to the steps (A5)-(A8) the same as the steps (S2)-(S5) of the BFK protocol \cite{BFK:UBQC:FOCS09}. In addition, there are some verification methods such as that in Ref. \cite{fitzsimons2017unconditionally} to make the BFK protocol to be verifiable and thus the proposed protocol also can be adapted to be a verifiable one. For example, the trap qubits used in step (A2) can disentangle them with the resource state and thus can be used for verifying the computation as Ref. \cite{fitzsimons2017unconditionally}.

\section{Comparisons among three typical BQC models and similar BQC protocols}
Before the new BQC model is proposed, there mainly exist two models of BQC. One is that the client just needs the capability to prepare single rotated qubits
initiated by Broadbent, Fitzsimons, and Kashefi \cite{BFK:UBQC:FOCS09}, and the other is that the user only needs perform single-qubit measurements first given by Morimae \cite{MF:BQC:PRA13}. In this paper, we propose a new BQC model that the client only requires the ability to implement a very limited set of single-qubit gates. We make comparisons among them mainly from the ability that a client requires, the ability that a server needs, and the interaction method as shown in Table \ref{tab:1}. Three models are suitable for practical implementations in different systems and all are necessarily to be considered. For example, to measure a single-qubit state is easier than to prepare a single-qubit state in optical system \cite{MF:BQC:PRA13}, and to perform single-qubit gates are more exact than to prepare or measure single-qubit states in superconducting systems \cite{barends2014superconducting}. But the proposed BQC protocol need two-way communication and thus it become less efficient since they need more particles to lower the influence of particles loss during transmission or reduce the transmission distance to be a half of that in the other two models.
\begin{table*}
\label{tb:1}
\newcommand{\tabincell}[2]{\begin{tabular}{@{}#1@{}}#2\end{tabular}}
\caption{Comparisons among the BFK protocol, Morimae's protocol, and the proposed protocol}
\label{tab:1}       
\resizebox{\textwidth}{12mm}{
\begin{tabular}{ccccc}
\hline\noalign{\smallskip}
    & The BFK protocol \cite{BFK:UBQC:FOCS09} & Tomorial's protocol \cite{MF:BQC:PRA13} & The proposed protocol \\
\hline\noalign{\smallskip}
Model types & Model of preparing single-qubit states & Model of measuring single qubits & Model of performing single-qubit gates \\
\hline\noalign{\smallskip}
Suitable physical systems & Semiconducting systems & Optical systems & Trapped ions and superconducting systems  \\
\noalign{\smallskip}\hline\noalign{\smallskip}
Client's quantum power         & \tabincell{c}{Generate single-qubit rotated states } & Perform single-qubit measurements & Perform gates $H$ and $\sigma_z^{1/4}$ \\
\hline\noalign{\smallskip}
Server's quantum power & \tabincell{c}{Full quantum} & Full quantum &  Full quantum  \\
\hline\noalign{\smallskip}
Interaction method & One-way (From Alice to Bob) & One-way (From Bob to Alice) & Two-way\\
\hline\noalign{\smallskip}
\end{tabular}}
\end{table*}

In addition, there exist some similar BQC protocols where a client needs limited quantum power including both performing single-qubit gates and other capabilities. Childs proposed a BQC protocol where the user Alice is required to implement Pauli $X$ gate, Pauli $Z$ gate, and the SWAP gate, own quantum memory, and prepare states $|0\rangle$ \cite{Childs:BQC:QIC05}. Ferracin et al. proposed a mesothetic verification protocol which can be made blind via an increase in circuit depth and the verifier Alice in it requires the ability to execute single-qubit gates and an $n$-qubit memory \cite{ferracin2019accrediting}. Comparisons among these two protocols and the proposed BQC protocol suitable for the given model from the ability that a client requires, the ability that a server needs, and the interaction method as shown in table \ref{tab:2}.

\begin{table*}
\label{tb:2}
\newcommand{\tabincell}[2]{\begin{tabular}{@{}#1@{}}#2\end{tabular}}
\caption{Comparisons among Childs's protocol, Ferracin et al's protocol, and the proposed protocol}
\label{tab:2}       
\resizebox{\textwidth}{12mm}{
\begin{tabular}{ccccc}
\hline\noalign{\smallskip}
    & Childs's protocol \cite{Childs:BQC:QIC05} & Ferracin et al's protocol protocol \cite{ferracin2019accrediting} & The proposed protocol \\
\noalign{\smallskip}\hline\noalign{\smallskip}
Client's quantum power         & \tabincell{c}{Perform gates Pauli $X$, Pauli $Z$, and SWAP \\own quantum memory \\prepare states $|0\rangle$} &  \tabincell{c}{Perform a set of single-qubit gates in the target circuit\\own quantum memory} & Perform gates $H$ and $\sigma_z^{1/4}$  \\
\hline\noalign{\smallskip}
Server's quantum power & \tabincell{c}{Full quantum} & Full quantum &  Full quantum \\
\hline\noalign{\smallskip}
Interaction method & One-way (From Alice to Bob) & Two-way & Two-way \\
\hline\noalign{\smallskip}
\end{tabular}}
\end{table*}

\section{Conclusion and discussion}

In this work, we have proposed a new BQC model in which the client only needs to do a limited set of single-qubit gates. It may be more suitable for practical implementation compared with the model of preparing states and that of making measurements due to that in some experimental setups such as trapped ions \cite{harty2016high} and superconducting systems \cite{barends2014superconducting}, the single-qubit gates are the most accurate operation \cite{ferracin2019accrediting}. In addition, it is very flexible since it can adapt to various clients owning different devices to implement some gates and thus enlarges the application of BQC. For example, some clients may have the ability to implement the gates in a set $\{H, \sigma_z^{1/4} \}$, but the others may only be able to perform the gates in another set $\{X, \sigma_z^{1/4} \}$. They all can finish the universal quantum computation with the help of a powerful quantum server.

However, it is not difficult to find whether a client can perform the gate $\sigma_z^{1/4}$ is very important in the proposed BQC protocols. Childs once showed that a client should initiate a two-round protocol with a quantum server to achieve it and the client needs the ability to generate quantum states and perform the SWAP gate, the Pauli $Z$ gate, and the Pauli $X$ gate \cite{Childs:BQC:QIC05}. So whether $\sigma_z^{1/4}$ can be substituted by other single-qubit gates which are easier to be implemented in experiments is an interesting question. In addition, we mainly focus on proposing a new BQC model where users only needs do single-qubit gates and just have constructed a simple BQC protocol based on the BFK protocol \cite{BFK:UBQC:FOCS09,fitzsimons2017unconditionally} as an example in such a model to show its feasibility. The two-round communication made the efficiency of the designed BQC protocol low and thus how to design efficient BQC protocols in the proposed model deserves further study.

\ack{We would like to thank Samuele Ferracin and Adrian Parra-Rodriguez who bring to our attention several interesting and related articles. This work was supported by the Joint Funds of the National Natural Science Foundation of China and China General Technology Research Institute (Grant No. U1736113), the Key Project of Hunan Province Education Department (Grant No. 20A471), Hunan Provincial Natural Science Foundation of China (Grant No. 2018JJ2403), and National Natural Science Foundation of China (Grant No. 61902132)}

\end{document}